\documentclass[pra,twocolumn,showpacs,superscriptaddress]{revtex4}
\usepackage{amsmath}
\usepackage{amssymb}
\usepackage{amsfonts}
\usepackage{graphicx}
\usepackage{bm}
\usepackage{enumerate}
\usepackage{color}
\usepackage{amsthm}
\renewcommand{\vec}[1]{{\bm{#1}}}

\begin{document}

\title{Complementarity and uncertainty relations for matter wave interferometry}
\author{Karl-Peter~Marzlin}
\affiliation{Department of Physics, St. Francis Xavier University,
  Antigonish, Nova Scotia, B2G 2W5, Canada}
\affiliation{Institute for Quantum Information Science,
        University of Calgary, Calgary, Alberta T2N 1N4, Canada}
\author{Barry C. Sanders}
\affiliation{Institute for Quantum Information Science,
        University of Calgary, Calgary, Alberta T2N 1N4, Canada}
\author{Peter L. Knight}
\affiliation{Blackett Laboratory, Imperial College, London SW7 2AZ, U.K.}

\begin{abstract}
We establish a rigorous quantitative connection between (i)~the interferometric duality relation 
for which-way information and fringe visibility and (ii)~Heisenberg's uncertainty relation 
for position and modular momentum. 
We apply our theory to atom interferometry, wherein spontaneously emitted photons 
provide which way information, and unambiguously resolve the challenge posed
by the metamaterial `perfect lens' to complementarity 
and to the Heisenberg-Bohr interpretation of the Heisenberg microscope thought
experiment.  
\end{abstract}

\maketitle

\section{Introduction}
Complementarity is at the heart of quantum mechanics and is operationally explored via
interferometry, specifically the quantitative trade-off between which path information~$W$ (for 
`which way') and visibility~$V$ (sharpness of
fringes)~\cite{San89,Scu91,Jae95,Eng96,Eng00,Wis97,Due00,Aha04}, which is a special case
of the information-disturbance trade-off~\cite{Woo79,Sac06}. 
An alternative view of complementarity is provided by the uncertainty
relations for position~$\vec{x}$ and momentum~$\vec{p}$, 
analyzed by Heisenberg in the context of $\gamma$-ray
microscopy~\cite{Hei29} to infer 
an electron's position at the expense of recoil due to 
collision with the short-wavelength photon. 

During the last two decades the question of how inevitable the recoil
is when $W$ is measured has received much attention.
In 1991, Scully {\em et al.}~\cite{Scu91} proposed a which-way measurement scheme that 
essentially would transfer no momentum to the particle, but
later Storey {\em et al.}~\cite{storey95} proved a general
theorem showing that any measurement of $W$ causes a momentum transfer
at least of order $\hbar/s$, with $s$ the spatial resolution of the
measurement scheme. In a careful analysis Wiseman {\em et al.}~\cite{Wis97} 
resolved this apparent contradiction using phase-space methods. They showed
that the momentum transfer in a measurement of $W$ can not always be understood
as a classical distribution of random recoils, but that under special 
circumstances \cite{Scu91} the momentum transfer is non-local and a genuine
quantum phenomenon.
In a recent experiment, Mir {\em et al.}~\cite{Mir07} have addressed this
using weak measurements \cite{Aha88,Dur98,Wis03}
to determine the momentum transfer 
in a photonic interferometer.

Since 1980 substantial progress has also been made
on the
quantitative analysis of $W$ and $V$, culminating in the duality relation
\cite{San89,Jae95,Eng96,Eng00} 
\begin{equation}  
        W^2 + V^2 \leq 1,
 \label{eq:WVineq}
\end{equation} 
which demonstrates the complementary nature of $W$ and $V$.
The equivalence between uncertainty relations, 
in particular the uncertainty relation between position
and momentum,
\begin{equation} 
  \Delta x \, \Delta p \geq \frac{\hbar}{2} \; ,
\end{equation} 
and the duality
relation has been the subject of debate. The claim that
they are logically independent \cite{Eng96} has been put into
question by D\"urr and Rempe \cite{Due00} who related the duality relation
to uncertainty relations between {\em Pauli matrices} for two-level
systems. Busch {\em et al.}~\cite{Bus06,Bus07} have presented a profound
analysis of the Mach-Zehnder interferometer and showed that
duality relations for the trade-off between partial path 
determinations and reduced-visibility interference observations
are expressible as uncertainty relations.
However, the common assumption that complementarity of $W$ and $V$ is closely
related to the uncertainty relation between {\em position and momentum} has not been
proven yet, and while Wiseman {\em et al.}~\cite{Wis97} beautifully
analyse the nature of the momentum transfer in measurements of $W$, they do
not investigate its relation to the duality relation.
Here we provide a quantitative relation between both concepts by
showing that the duality relation can be used to derive an
uncertainty relation between position $x$ and  
modular momentum~$\tilde{p}$.

A second challenge to interferometric complementarity suggests that
superresolution~\cite{Roy78} or perfect resolution~\cite{Ber03} 
from metamaterial `perfect lenses'~\cite{Pen00,Smi04} 
is not easily reconciled with complementarity and interferometry 
because the Bohr-Heisenberg interpretation of Heisenberg's
$\gamma$-ray microscope~\cite{Boh29} links the uncertainty relation
to the optical diffraction limit. Whereas Roychoudhuri expressed
doubts about this interpretation, his argument is qualitative~\cite{Roy78}; in contrast
we rigorously and quantitatively resolve this challenge by showing that the perfect lens simply 
provides an extremal point in the duality relation for atom interferometry.

\section{Atom interferometry}
In atom interferometry~$W$ quantifies to what extent it can be
predicted through which of the two paths an atom will travel.
Visibility $V$ is a measure for the contrast of the interference pattern.
Both are usually taken to be a number between 0 and 1. If $W$ assumes
the maximum value 1, the atom passes with certainty through only
one of the two paths. Obviously, this would prohibit any interference
phenomena between the two paths so $V$ should be zero in this case.
On the other hand, if $W=0$ the probabilies for the atom to pass through
either path are equal. If the atom is prepared in a coherent superposition
of both paths then $V$ can be maximal. However, if the atom is prepared
in an equally weighted mixture to pass through either path, then interference
phenomena would still be impossible so that $W=V=0$.

To establish a connection between duality and uncertainty
of position and momentum we obviously have to quantize the atomic 
center-of-mass (CoM) motion. An atom then has internal (electronic)
and CoM degrees of freedom, and it is the latter which will be in the
focus of our attention. If an atom is prepared in 
a (CoM-) state localized around $\vec{x}=\vec{0}$,
corresponding to one arm of the interferometer, it is described by a 
normalized wave function $\phi\left(\vec{x}\right)$.  
A wavepacket that has the same shape but is localized around
$\vec{x}=\vec{a}$ is given by 
\begin{equation}
\label{eq:T}
        \phi\left(\vec{x}-\vec{a}\right)=\hat{T}_{\vec{a}}\phi\left(\vec{x}\right),\;
        \hat{T}_{\vec{a}} = \exp\left(-i\vec{a}\cdot\hat{\vec{p}} /\hbar\right)
\end{equation}
where~$\hat{T}_{\vec{a}}$ is a shift operator and $\hat{\vec{p}}$ the vector momentum operator. 

\begin{figure}
\includegraphics[width=8cm]{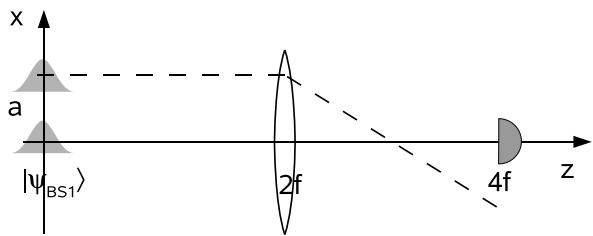} 
\caption{\label{microscSketch} 
  Model for the Heisenberg microscope. The two atomic wave functions
  are located close to the origin and
  are separated by a small distance $a$ in the x-direction. The spontaneously
  emitted light that is collected in the detector propagates paraxially
  along the $z$-axis. The lens is located at $z=2f$ and the detector at
  $z=4f$.} 
\end{figure}

In this paper we will consider the case that the process of splitting
the atomic beam does not distort the shape of the beam so that
the wavepacket that describes the second arm of the interferometer can
be described by Eq.~(\ref{eq:T}).
The atomic CoM wave function
is initially prepared in the state 
$\phi\left(\vec{x}\right) = \langle x|\phi \rangle $.
A generic atom beam splitter consists of a grating \cite{Car91} or employs
light forces \cite{Kas92}. When the process of splitting the beam is
completed the atomic state after the first beam splitter is given by
\begin{equation} 
  |\psi_\text{BS1} (\theta) \rangle =  \frac{1}{n_\text{BS1}} 
  \left( |\phi \rangle + e^{i\theta}
   \hat{T}_{\vec{a}} |\phi \rangle  \right) 
  = U_\text{BS}(\theta)\,|  \phi \rangle  \; ,
\label{BS1eq}\end{equation} 
which corresponds to a superposition of the wavepackets at two locations. 
Here $n_\text{BS1}$ is a normalization that ensures 
$ \langle \psi_\text{BS1} (\theta) |\psi_\text{BS1} (\theta) \rangle=1$.
The state $|\psi_\text{BS1} (\theta) \rangle$ corresponds to the two
localized wavepackets on the left-hand side of Fig.~\ref{microscSketch}.
We therefore can model the beam splitting process by a Householder 
reflection~\cite{lehoucq96:_comput} $U_\text{BS}(\theta)$,
which is a unitary transformation that maps a given state $|\phi \rangle $
to a given, non-orthogonal state $|\psi_\text{BS1} (\theta) \rangle$. 
The Householder reflection is not uniquely defined by these two states;
one convenient form is given by
\begin{equation} 
  U_\text{BS}(\theta) \equiv \frac{
  (|\phi \rangle + |\psi_\text{BS1}(\theta) \rangle)\,
  (\langle \phi | + \langle \psi_\text{BS1}(\theta)|)
  }{1+ \langle  \psi_\text{BS1}(\theta) |\phi \rangle} 
  - \hat{\vec{ 1}} \; .
\end{equation} 
In Eq.~(\ref{BS1eq}) we included an arbitrary relative phase shift $\theta$ 
between the two beams in the beam splitting process. In an experiment
it would be generated by a phase shifter in one beam right after the
beam splitter. Varying $\theta$ will enable us to explore the fringe
pattern $f(\theta)$ of the interferometer, which will be necessary
to collect information on~$V$ (see below). This is the reason why we highlight
the dependence of $| \psi_\text{BS1}(\theta)\rangle $ on $\theta$.
Throughout the paper we assume the large mass limit so the wave functions 
are effectively immutable during beam splitting and during a which way
detection. This assumption is central for our analysis of complementarity.

$W$ is obtained by performing
a generalized position measurement~\cite{Sto94,Wis97} on the split
atomic beam. Loosely speaking this is a measurement that can determine
the position only up to a certain accuracy; i.e., each possible
measurement outcome has some uncertainty.
Mathematically a generalized position measurement 
is described by a set of functions
${\cal D}_\alpha(\vec{x})$, where $\alpha$ runs over some index set. 
These functions form a partition of unity of the form
\begin{equation} 
  \sum_\alpha |{\cal D}_\alpha(\vec{x})|^2 =1 \quad \forall \vec{x} 
\end{equation} 
After a generalized position measurement has produced the result $\alpha$,
the atomic state will be modified according to
\begin{equation} 
  \psi(\vec{x}) \rightarrow  n_\alpha {\cal D}_\alpha(\vec{x}) \psi(\vec{x})\; ,
\end{equation} 
where $n_\alpha$ is a normalization factor. In the following we will only
be interested in the state after a generalized position measurement has
generated a specific result; we therefore will drop the index $\alpha$
and denote with ${\cal D}(\vec{x})$ that function which corresponds to
this specific result. 
In Sec.~\ref{Sec:PerfectLensPlusComplementarity} we will show 
that detection of a spontaneously emitted photon corresponds 
to such a generalized position measurement. In the set-up
shown in Fig.~\ref{microscSketch} this measurement is performed
by collecting the emitted light using a lens at position $z=2f$
and detecting the light at $z=4f$, where $f$ is the focal length
of the lens.

For the atom interferometer under consideration,
the postselected state after the position measurement is given by
\begin{eqnarray} 
  \psi_{{\cal D}}\left(\vec{x}\right) &=&
  \frac{1}{\sqrt{n(\theta)}} 
  \left (
  \mathcal{D}(\vec{x})\phi\left(\vec{x}\right)
  + e^{i\theta}  \mathcal{D}(\vec{x})\hat{T}_{\vec{a}}
  \phi\left(\vec{x}\right) 
  \right )
\label{eq:detect2} \\
\\
       n(\theta) &\equiv& (n_0 + n_1) (1+ \text{Re}\left(e^{i\theta} r\right))
\\
       r &\equiv&  \frac{ 2 \langle \phi | {\cal D}^\dagger(\hat{\vec{x}})
         {\cal D}(\hat{\vec{x}})\, \hat{T}_{\vec{a}} | \phi \rangle }{n_0+n_1}\; .
\\  
  n_0 &\equiv& \langle \phi |{\cal D}^\dagger(\hat{\vec{x}}) {\cal D}(\hat{\vec{x}})|\phi \rangle 
\\  
  n_1 &\equiv& \langle \phi |\hat{T}_{\vec{a}}^\dagger  {\cal D}^\dagger(\hat{\vec{x}}) 
  {\cal D}(\hat{\vec{x}})\hat{T}_{\vec{a}} |\phi \rangle 
\end{eqnarray} 
The factor $n(\theta)$
ensures that $\psi_{{\cal D}} (\vec{x})$ is
normalized, and the parameter $r$ is proportional to
the overlap between the two paths of the interferometer.
The parameters $n_0$ and $n_1$ are proportional to the populations in each path 
of the interferometer.
\section{Complementarity}
$W$ quantifies the difference of the probabilities to find the atom in 
the two interferometric paths 
$|\phi \rangle $ or $\hat{T}_{\vec{a}} |\phi \rangle $.
In our case we collect which way information by performing
a generalized position measurement; we therefore have to calculate
$W$ for the atomic state after this measurement has been performed.
For non-overlapping interferometer paths $W$ can simply be defined as
the difference of the probabilities to find the atom in either path.
If the paths do overlap, $W$ relates to the distinguishability of the two paths.

We wish to employ a conclusive protocol for identifying which of
two non-orthogonal states best describes the preparation of the
system. If the state is found to be in one of the two states, then
we can be certain this is the prepared state, but the price is
that a third measurement must be allowed: the null measurement. If
the result is a null measurement, then we are completely uncertain
about which state was prepared. The optimal positive
operator-valued measure (POVM) for a conclusive protocol for two
non-orthogonal states is given by the rank-three set of operators~\cite{POVM}
\begin{eqnarray} 
   \hat{P}_0 &=& \frac{\openone-\left|\phi \rangle \langle
   \phi \right|}{1 +|\langle \phi|\hat{T}_{\vec{a}} | \phi \rangle |} 
\\
   \hat{P}_1 &=& \frac{\openone- \hat{T}_{\vec{a}} \left|\phi \rangle \langle
   \phi \right|\hat{T}_{\vec{a}}^\dagger }{1 +|\langle \phi|\hat{T}_{\vec{a}} | \phi \rangle |} 
\\   \hat{P}_2 &=& \openone- \hat{P}_0 -\hat{P}_1 \; ,
\end{eqnarray} 
where $\hat{P}_2$ corresponds to the null measurement.
Employing this POVM
and using that the probabilities to be in state $|\psi_i \rangle$ are given by
$P_i = \langle \psi_{{\cal D}}| \hat{P}_i |\psi_{{\cal D}} \rangle $, we obtain
\begin{eqnarray}  
    \tilde{W}(\theta) &=& \left| P_0-P_1\right|
\nonumber \\ &=&
  \frac{n_0+n_1}{(1+|\langle \phi|\hat{T}_{\vec{a}} | \phi \rangle |
    ) n(\theta)}
  \Big ( |z_0|^2  - |z_1|^2 + |z_2|^2
\nonumber \\ & &
   - |z_3|^2  + \big ( e^{i \theta} (z_0^* z_2
  -z_3^* z_1) +\text{c.c.} \big) \Big )\; ,
\label{Wnewdef0}
\end{eqnarray} 
where we have introduced the complex numbers
\begin{eqnarray} 
  z_0 &\equiv & \frac{ \langle \phi| {\cal D}(\hat{\vec{ x}}) |\phi \rangle}{ \sqrt{n_0+n_1}}
\\
  z_1 &\equiv & \frac{ \langle \phi|\hat{T}_{\vec{a}}^\dagger  {\cal D}(\hat{\vec{ x}})
  \hat{T}_{\vec{a}} |\phi \rangle }{\sqrt{n_0+n_1}}
\\
  z_2 &\equiv & \frac{ \langle \phi| {\cal D}(\hat{\vec{ x}}) \hat{T}_{\vec{a}} |\phi
  \rangle}{ \sqrt{n_0+n_1}}
\\
  z_3 &\equiv & \frac{ \langle \phi|\hat{T}_{\vec{a}}^\dagger  
  {\cal D}(\hat{\vec{ x}}) |\phi \rangle }{ \sqrt{n_0+n_1}}
 \; .
\end{eqnarray} 
For overlapping wavepackets $\tilde{W}(\theta)$ depends on the interference phase
$\theta$ because constructive and destructive interference in the overlap
region can decrease and increase the distinguishablility, respectively. 
To achieve a measure for which way information that is independent of the phase
we define the which way information $W$ as the mean of $\tilde{W}(\theta)$, 
\begin{eqnarray} 
  W &=& \frac{1}{2} \left ( \tilde{W}(\theta_\text{max})+ \tilde{W}(\theta_\text{min}) \right )
\nonumber \\ &=&
  \frac{1}{(1+|\langle \phi|\hat{T}_{\vec{a}} | \phi \rangle |
    ) (1-|r|^2)}
  \Big ( |z_0|^2 - |z_1|^2 + |z_2|^2 
\nonumber \\ & &
   - |z_3|^2  - \big ( r^* (z_0^* z_2  -z_3^* z_1) +\text{c.c.} \big) \Big )\; .
\label{Wnewdef1}\end{eqnarray} 
If the two wavepackets are non-overlapping
$\tilde{W}(\theta)$ and $W$ agree. We then have 
$\langle \phi|\hat{T}_{\vec{a}} | \phi \rangle = r = z_2 = z_3 = 0$ and $W$ reduces to
\begin{equation} 
  W_\text{no} =  |z_0|^2 -|z_1|^2 \; ,
\end{equation} 
which corresponds to the population difference in both arms of the interferometer
after the generalized position measurement has been performed.
For perfect
overlap one has $\langle \phi|\hat{T}_{\vec{a}} | \phi \rangle = r = 1$ and $z_0=z_1=z_2=z_3$.
This results in vanishing which way information $W=0$, 
which is a consequence of the two beams being indistinguishable.

The fringe visibility $V$ is obtained by recombining the 
two atomic beams (which is described by a unitary 
transformation $U$)  and to equate $V$
with contrast. The latter corresponds to the normalized difference
\begin{equation} 
  V = \frac{f(\theta_\text{max})-f(\theta_\text{min})}{
  f(\theta_\text{max})+f(\theta_\text{min})}
\label{vSqrd} \end{equation} 
between the maximum and minimum of the fringe pattern  $f(\theta)$.
If the processes of measuring which way information, recombining the beam, and detecting the
atoms do not alter the shape of the atomic wavepacket, one may describe 
the interferometer with just two states (one for each beam) \cite{Eng96} .
The fringe pattern  $f(\theta)$
can then be observed by measuring the overlap of the incoming atomic state $|\psi_\text{in}\rangle$
with the output of the interferometer, which corresponds to a measurement of the observable 
$|\psi_\text{in}\rangle\langle \psi_\text{in}|$.
In our case the which way measurement may in general change the wavepacket, but a 
straightforward generalization of the previous observable is the overlap
between the recombined state $U_\text{BS}^\dagger(0)   | \psi_b(\theta) \rangle $
and the input state $|\phi \rangle$, so that
\begin{eqnarray} 
  f(\theta) &=& | \langle \phi|U_\text{BS}^\dagger(0)   | \psi_b(\theta) \rangle |^2 \; .
\end{eqnarray}  
This yields 
\begin{eqnarray} 
  V^2 &=&1- (1-|r|^2) \left (|z_0+z_3|^2 -|z_1+z_2|^2 \right )^2
  \Big ( |z_0+z_3|^2 
\nonumber \\ & & 
  +|z_1+z_2|^2 -\big (  
  r (z_0+z_3)(z_1^*+z_2^*) +\text{c.c}
  \big )
  \Big )^{-2}
\end{eqnarray} 
In the limit of non-overlapping wavepackets visibility reduces to
\begin{equation} 
  V_\text{no} = \frac{2 |z_0| \, |z_1| }{|z_0|^2 + |z_1|^2}\; .
\end{equation} 
For completely overlapping wavepackets we find $V=0$, which is again a consequence
of the indistinguishability of the paths. 

The duality relation (\ref{eq:WVineq}),
which conveys that there is an informational trade-off between which path information
and visibility~\cite{Woo79},
can easily be verified in the case of non-overlapping
wavepackets. Using the Cauchy-Schwartz inequality 
\begin{equation} 
  | \langle \psi_1 | \psi_2 \rangle |^2 \leq \langle \psi_1 | \psi_1 \rangle
  \langle \psi_2 | \psi_2 \rangle \quad \forall |\psi_1 \rangle , 
  |\psi_2 \rangle \in {\cal H}\; ,
\label{CSineq} \end{equation} 
one finds
$  |z_0|^2 \leq n_0/(n_0+n_1) $ and $  |z_1|^2 \leq n_1/(n_0+n_1) $ so that
$|z_0|^2+ |z_1|^2\leq 1$. This condition guarantees that 
$V_\text{no}^2 + W_\text{no}^2 \leq 1$. Even though the state 
$|\psi_b \rangle $ is pure, the duality relation is exactly fulfilled
only in the special cases that (i) $|z_0|^2+ |z_1|^2 = 1$ and (ii) $|z_0|=|z_1|$.
In case~(i) the detector function ${\cal D}$ has perfect overlap with 
$|\psi_b \rangle $. This implies that $\langle \psi_b |\hat{P}_2 |\psi_b
\rangle =0$ so that the detector provides complete knowledge about
complementarity. Case~(ii) corresponds to the situation that the
interferometer is perfectly balanced 
(the atom travels through both paths with equal probability) 
even after the position measurement.
Hence $W_\text{no}=0$ and, because the contrast of the fringes is not affected
by a non-perfect overlap of the detector function, $V_\text{no}=1$. 

\begin{figure} 
\includegraphics[width=5.5cm]{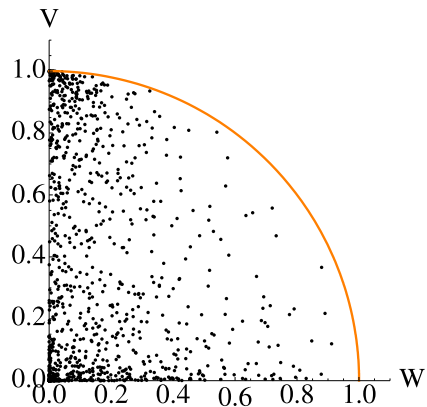} 
\caption{\label{randomStates} 
  (Color online) 
  Verification of the duality relation for a set of 1000 random states. The duality
  relation is valid for states inside the orange circle, see text for details.} 
\end{figure}

It seems obvious that the duality relation should also be fulfilled for
overlapping states because any overlap should decrease the distinguishability
between the two interferometer arms and thus reduce $W$ and $V$.
However, a general proof of this conjecture is surprisingly difficult
\footnote{
It seems impossible to prove Eq.~(\ref{eq:WVineq}) analytically for
arbitrary atomic states. Instead, two other approaches can be used
to test its validity. (i) a numerical evaluation of
Eq.~(\ref{eq:WVineq}) for a set of randomly generated values for
the parameters $n_i, r, z_i$. However, not all possible
real or complex values for $n_i, r, z_i$ correspond to a state. 
For instance, from the definition of $W$ we know that $W<1$ for
all states, but it is easy to see that in Eq.~(\ref{Wnewdef1})
$W\rightarrow \infty$ for $r\rightarrow 1$ and $z_i$ fixed.  
To generate only physical parameter values we have therefore constrained the 
random values by a set of 20 inequalities that we derived using the Cauchy-Schwartz 
inequality and general relations for the overlap between two given states. 
A typical example of one of the 20 inequalities
would be $|\langle \phi|{\cal D}^\dagger {\cal D}(\hat{T}_{\vec{a}}
  - \langle \hat{T}_{\vec{a}} \rangle) | \phi \rangle|^2 \leq 
n_0 (1-|\langle \hat{T}_{\vec{a}} \rangle|^2)$. However, even this large
number of constraints did not exclude certain unphysical values for the parameters,
and thus this approach did not help to verify Eq.~(\ref{eq:WVineq}).
(ii) A second approach to verify Eq.~(\ref{eq:WVineq}) is to numerically evaluate $W$ and $V$ for a random
set of quantum states. This is the approach described in the text.
}. Instead, we have verified numerically that the duality relation holds
for a sample of 100,000 random Gaussian states, where $\hat{T}_{\vec{a}} \phi(x)$ 
takes the form $\exp(- (x-x_0)^2/w^2 + i k x)$. The detector function
${\cal D}(x)$ takes a similar form but with different parameters
$x_0, w$, and $k$ that were chosen randomly for both $\phi(x)$ and
${\cal D}(x)$ and were allowed to vary between -4 and 4 in units of the
width of the width of the initial Gaussian state $\phi$. 
The results for a
sub-sample of 1000 random states are shown in Fig.~\ref{randomStates}. 
We found no violation of Eq.~(\ref{eq:WVineq}).

\section{Complementarity and uncertainty relations}
In this section we offer a new perspective on the ongoing debate
whether the duality relation~(\ref{eq:WVineq}) 
is logically independent of Heisenberg's uncertainty 
relation~\cite{Eng96,Dur98,Kni98} or not~\cite{Due00}.
To address this conundrum we consider the special case of non-overlapping
wavepackets. 
For simplicity we restrict our considerations to the spatial component $x$ that
is parallel to the separation vector $\vec{a}$ between the two interferometer paths. 
We assume that $\phi(x)$ is a wavepacket of arbitrary shape that is centered around 
the origin, with a width that is small compared to the separation $a\equiv |\vec{a}|$ 
between the two beams. In this case the position uncertainty induced by the
finite width of $\phi(x)$ is generally negligible as compared to that induced by the
superposition of the two interferometer paths $\phi(x)$ and $\phi(x-a)$.
We then can make the approximation 
$
  \langle \phi| (\hat{x}+ a)^n |\phi
  \rangle \approx  a^n
$ so that
$\langle \psi_{{\cal D}}| \hat{x}^n|\psi_{{\cal D}} \rangle \approx a^n n_1/(n_0+n_1)$.
The position uncertainty in state $| \psi_{{\cal D}}\rangle$ then
simplifies to
\begin{equation} 
  \Delta \vec{x}^2 \approx   a^2 \frac{n_0 n_1}{(n_0+n_1)^2} \; .
\end{equation} 
Because which way information quantifies the probabilities for an atom to take one
of the two interferometer paths one would generally expect a close relation between $W$
and $\Delta x$. For instance, if $W=1$ one knows with certainty that the atom took
one of the two paths so that $\Delta x$ should be comparable to the width of the
wavepacket $\phi(x)$. On the other hand, if $W=0$ then it is uncertain which
path the atom takes. Then $\Delta x$ should be of the order of the path
separation $\vec{a}$ which may be much larger than the width of the wavepacket.
However, the argument above does not take the quality of the position measurement into account.
If we can make the same approximations in the evaluation of ${\cal D}(\hat{x})$
as in that of $\Delta x$, then a Taylor
expansion of the detector function yields $\langle \phi| {\cal D}(\hat{x}) |\phi
\rangle \approx {\cal D}(0)$, which results in
\begin{equation} 
  \frac{\Delta x^2}{a^2} \approx \frac{1}{4}(1-W^2) \approx
  \frac{|{\cal D}(0)|^2 |{\cal D}(a)|^2 }{(|{\cal D}(0)|^2 +|{\cal D}(a)|^2
    )^2}\; .
\label{dxwrelation}\end{equation} 

\begin{figure}
\includegraphics[width=8cm]{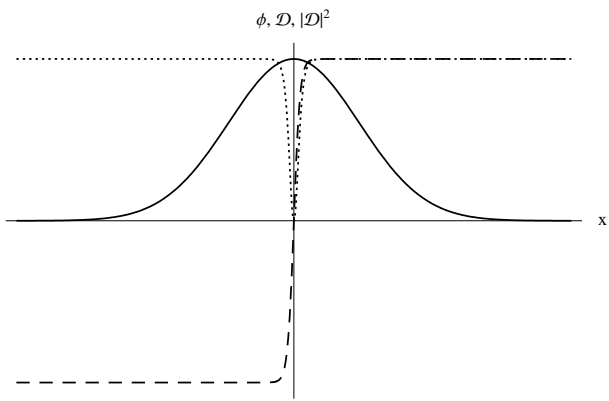} 
\caption{\label{Wdxscheme} 
A combination of a wavepacket (solid line) and a detector function (dashed
line) for which there is no relation between which way information and
position uncertainty. The dotted line corresponds to $|{\cal D}|^2$.} 
\end{figure}

However, this exact relationship between $\Delta x$ and $W$ is only
valid if $\langle \phi| {\cal D}(\hat{x}) |\phi \rangle \approx {\cal D}(0)$,
i.e., if the detector function ${\cal D}( x)$ varies little over the the
extent of the wavepacket $\phi(x)$. The example presented in
Fig.~\ref{Wdxscheme} demonstrates that a rapid variation of 
${\cal D}(x)$ can affect relation (\ref{dxwrelation}).
In this case a symmetric wavepacket is combined with an 
antisymmetric detector function
so that  $\langle \phi| {\cal D}(\hat{x}) |\phi \rangle =0$ and consequently
$z_0=0$. On the other
hand, $|{\cal D}(x)|^2$ is close to unity almost everywhere so that
$n_0 = \langle \phi| |{\cal D}(\hat{x})|^2 |\phi \rangle \approx 1$.
If we assume that ${\cal D}(x)=0$ around $x=a$ then
$n_1=z_1=0$ so that $\Delta x \approx W \approx 0$. Hence a detector only
gathers which way information if the detector function ${\cal D}$ is suitable.

Even in the case of a general detector function ${\cal D}(x)$
it is possible to establish an inequality that relates position uncertainty
and which way information. Using the Cauchy-Schwartz inequality (\ref{CSineq}) 
in the form $  |z_i|^2 \leq n_i/(n_0+n_1)\; , i=0,1 $ one easily finds
\begin{equation} 
  \frac{1}{2}(1-W^2) \geq \frac{\Delta x^2}{a^2} \; ,
\label{dxwInequality}\end{equation} 
which reflects the fact that 
\begin{eqnarray} 
  \Delta x^2 &=&
  \int d^3x \, \psi_{{\cal D}}^*(x)\, (x-\langle x\rangle  )^2 
  \psi_{{\cal D}}(x)
\nonumber \\ &=&
    \int d^3x \, 
  (x-\langle x\rangle  )^2  |{\cal D}(x)|^2
  | \psi_{\theta}(x)|^2
\end{eqnarray} 
is not affected by the phase 
of ${\cal D}(x)$ and hence is less sensitive to rapid variations
of the detector function.

The connection between momentum uncertainty and visibility is more
subtle.
It is well known that
interference experiments do not directly reveal information
about the momentum but rather about the 
modular momentum~$\hat{T}_{\vec{a}}$~\cite{Aha04}.
On the other hand, $\hat{T}_{\vec{a}}$ of Eq.~(\ref{eq:T}) can also be 
associated with a ``phase observable'' $-a\cdot \hat{p} /\hbar$.
If we were able to measure this phase observable directly, then 
it would be possible to relate the duality relation directly to
the uncertainty of position and momentum. However, as with any phase
measurement in quantum mechanics, this is impossible, and we instead have
to consider the modular momentum. Holevo~\cite{Hol84} devised a method
of how to define a phase uncertainty for general observables of this form:
$\text{arg}\langle\hat{T}_{\vec{a}}\rangle$ 
corresponds to the mean phase of~$\hat{T}_{\vec{a}}$,
and the Holevo phase uncertainty is given by
\begin{equation} 
        \Delta_H =\left| \left\langle \hat{T}_{\vec{a}} \right\rangle \right|^{-2} -1
                       \;  \approx \; \Delta p^2/\hbar^2  \; .
\end{equation} 
The approximation applies for small deviations of $p$, which represents
the vector component of~$\vec{p}$ in the direction of separation~$\vec{a}$. 

An uncertainty relation between position and modular momentum
can be derived by adapting Heisenberg's derivation to 
a unitary operator $\hat{T}_{\vec{a}}$. For 
$\delta\hat{x}\equiv\hat{x}-\langle \hat{x} \rangle$ and
$|\psi_x \rangle \equiv \delta\hat{x} |\psi \rangle $,
$|\psi_T \rangle \equiv \hat{T}_{\vec{a}} |\psi \rangle$,
the Cauchy-Schwartz inequality (\ref{CSineq}) yields
\begin{equation} 
    |\langle \psi |(\delta\hat{x}) \hat{T}_{\vec{a}}| \psi\rangle |^2
   \leq \Delta x^2  \; ;
\end{equation} 
similarly
$| \langle \psi| \hat{T}_{\vec{a}} (\delta\hat{x})| \psi \rangle|^2$,
which results in
\begin{equation} 
  |\langle \psi|\, [\, \hat{T}_{\vec{a}},\delta\hat{x}] \,| \psi\rangle |/2
  \leq \Delta x \; .
\end{equation} 
On the other hand $[\hat{T}_{\vec{a}} , \hat{\vec{x}}] =- \vec{a} \,\hat{T}_{\vec{a}}$
so
\begin{equation} 
  \frac{ \Delta x}{a} \geq \frac{1}{2} |\langle \hat{T}_{\vec{a}} \rangle|
        = \frac{1}{2\sqrt{1+\Delta_H}}\; ,
\label{dxT}\end{equation} 
which establishes an uncertainty relation between position and
modular momentum. 

We now turn to the question whether uncertainty relation (\ref{dxT})
can be related to the duality relation (\ref{eq:WVineq}). For general
${\cal D}(\vec{x})$ only inequality (\ref{dxwInequality}) holds; 
we conjecture that in this case it is not possible to relate uncertainty 
and complementarity. The situation is different for suitable 
(i.e., slowly varying over the width of $\phi(x)$) 
detector functions
which fulfill 
$\langle \phi| {\cal D}(\hat{x}) |\phi \rangle \approx {\cal D}(0)$.
We then have $V\approx 2 |{\cal D}(0)|\, |{\cal D}(a)| 
/(|{\cal D}(0)|^2 +|{\cal D}(a)|^2 )$ and
\begin{eqnarray}  
  \left | \langle \psi_{{\cal D}}| \hat{T}_\vec{a} |\psi_{{\cal D}} \rangle \right |
  &=& 
  \left | \frac{e^{-i\theta}}{n_0+n_1} \langle \phi |\hat{T}_\vec{a}^\dagger \hat{{\cal D}}^\dagger
   \hat{T}_\vec{a} \hat{{\cal D}}|\phi \rangle \right |
\nonumber \\ &\approx &
  \left | \frac{{\cal D}^*(a)\, {\cal D}(0) }{(|{\cal D}(0)|^2 +|{\cal D}(a)|^2 )} 
 \right |
\nonumber \\ &=& \frac{V}{2} \; .  
\label{VTrelation}\end{eqnarray} 
Hence, for non-overlapping atomic beams and a suitable 
detector function
there is a direct relation between complementarity and the uncertainties of
position and modular momentum.
Inserting Eqs.~(\ref{dxwrelation}) and (\ref{VTrelation}) into the duality
relation (\ref{eq:WVineq}) immediately yields
\begin{equation} 
  \frac{\Delta x^2}{a^2} \geq |\langle \hat{T}_{\vec{a}} \rangle |^2 \; ,
\label{WVineq2}\end{equation} 
from which
the uncertainty relation (\ref{dxT}) between position and modular
momentum can be deduced. 
Therefore, for well-separated wavepackets
the duality relation appears stronger than
the Heisenberg uncertainty relation because the former can be used
to derive the latter.

\section{Perfect lens and complementarity}\label{Sec:PerfectLensPlusComplementarity}
Roychoudhuri~\cite{Roy78} and Berman~\cite{Ber03} 
have challenged the Heisenberg-Bohr explanation of complementarity
in the $\gamma$ ray microscope, which relates uncertainty to
the diffraction of the lenses that are used to collect the radiation emitted 
by the atom. 
They pointed out that within this interpretation
optical superresolution and diffraction-less metamaterial perfect lenses
would lead to a violation of the uncertainty principle.
Here we resolve this question by demonstrating that the detection of
light emitted by a two-level atom (2LA) in an atom interferometer corresponds to
a generalized position measurement. The quality of the lenses therefore
can only affect the amount of which way information that can be obtained,
but it cannot affect the duality relation (\ref{eq:WVineq}).
2LA interferometry and complementarity 
has previously been studied in Ref.~\cite{Wis97}, but this analysis did not
consider the perfect lens; here we
provide an alternative derivation that accommodates almost arbitrary 
arrangements of linear lossless dielectrics. We ignore the polarization of light
in our derivation because it will not substantially affect our results.

We consider the situation that 2LAs are excited immediately after the
beam has been split and then undergo spontaneous decay. 
As depicted in Fig.~\ref{microscSketch},
the spontaneously emitted photon is detected after passing through an array of
linear optical elements (which could include a perfect lens~\cite{Smi04}).
Just after excitation, the atomic state is
$\int  \text{d}^3\vec{x}\;\psi^{(0)}(\vec{x}) | \vec{x} \rangle \otimes|{\sf e}\rangle$
for $|{\sf e}\rangle$ the internal excited state.
Spontaneous emission over time scale~$1/\gamma$
returns the 2LA to its ground state~$|{\sf g}\rangle$. Here $\gamma$
is the decay rate of the atom in the presence of the dielectrics.
A crucial assumption for our derivation is that  $1/\gamma$ is short compared 
to the time scale $\tau_A$ during
which the atomic center-of-mass wavepacket changes significantly. This assumption
allows us to neglect the kinetic center-of-mass energy of the atoms
and should be valid for most situations. Exceptions would be atomic
ensembles very far from equilibrium, for which $\tau_A$ could be short,
or optical cavities of extremely high finesse for which $\gamma$ could
be significantly smaller than the natural atomic decay rate in free space.
The atomic Hamiltonian is then given by
\begin{equation} 
  \hat{H}_A = \hbar \omega_A |e \rangle \langle e|\; ,
\end{equation} 
with $\omega_A$ the resonance frequency of the 2LA.

Because the dielectrics are assumed to be linear and lossless, there is
a set of eigenmodes $E_n(\vec{x})$ with frequency $\omega_n$. For simplicity
we restrict our analysis to a discrete set of modes, but generalizing our
approach to a continuous set of modes should not affect the results.
The radiative Hamiltonian in the presence of dielectrics then takes the
general form
\begin{equation} 
  \hat{H}_R = \hbar \sum_n \omega_n \hat{a}^\dagger(n) \hat{a}(n) \; ,
\end{equation} 
where $\hat{a}(n)$ annihilates one photon in mode $E_n(\vec{x})$.
Implicitly we have assumed here that the dielectrics are time independent
over the time scale $1/\gamma$,
which is the case for almost all experiments except for very special
situations such as Faraday media driven by time varying external fields.
We describe the coupling between matter and radiation in electric-dipole
and rotating-wave approximation,
\begin{equation} 
  \hat{H}_\text{int} = -  d_{eg} |e \rangle \langle g|
   \sum_n  \hat{a}(n) E_n(\hat{x}) 
  + \text{H.c.}
\end{equation} 
Expanding the total state of the system as
\begin{eqnarray} 
  |\psi(t) \rangle &=&
  \int d^3x \Big ( \psi_e(\vec{x},t) |e \rangle \otimes | \vec{x} \rangle \otimes 
  |\text{vac} \rangle 
\nonumber \\ & &
  + \sum_n \psi_n (\vec{x},t) |g \rangle \otimes | \vec{x} \rangle \otimes 
  \hat{a}^\dagger(n)  |\text{vac} \rangle \Big ) \; ,
\end{eqnarray} 
with $|\text{vac} \rangle$ the radiative vacuum state, the Schr\"odinger 
equation can be cast into the form
\begin{eqnarray} 
  i \dot{\psi}_e(\vec{x}) &=& \omega_A \psi_e(\vec{x}) - 
  \frac{d_{eg}}{\hbar} \sum_n E_n(\vec{ x}) \, \psi_n(\vec{x})
\\
  i \dot{\psi}_n(\vec{x}) &=& \omega_n \psi_n(\vec{x}) -
  \frac{d_{eg}^*}{\hbar} \psi_e(\vec{x}) E_n^*(\vec{x})\; .
\end{eqnarray} 
Performing a Laplace transformation in the time domain allows us to find the solution as
\begin{eqnarray} 
  \tilde{\psi}_e(\vec{x},s) &=& \frac{ i \psi^{(0)}(\vec{x}) }{
  i s -\omega_A - \frac{|d_{eg}|^2}{\hbar^2} \sum_m
    \frac{|E_m(\vec{x})|^2}{is-\omega_m}
    }
\\
  \tilde{\psi}_n(\vec{x},s) &=& -\frac{d_{eg}^*}{\hbar}
  \frac{E_n^*(\vec{x})}{is-\omega_n}     \tilde{\psi}_e(\vec{x},s) \; ,
\end{eqnarray} 
where $\tilde{f}(s)$ denotes the Laplace transform of $f(t)$.
The solution in time domain can be expressed through the inverse Laplace transform 
\begin{equation} 
  \psi_n(\vec{x},t) = \frac{ 1}{2\pi i} \int_{{\cal C}} 
   \text{d}s\; \text{e}^{ts}  \tilde{\psi}_n(\vec{x},s)\; ,
\end{equation} 
with the path ${\cal C}$ being to the right of all poles and branch cuts.

This solution contains the photon dynamics in the presence of 
linear dielectrics. At time $t$ a detector is switched on to
register the emitted photon. We model the detector as a device that 
detects photons in a particular mode
characterized by the a specific superposition of annihilation operators 
$\hat{b} = \sum_n \eta^*(n) \hat{a}(n)$. 
The 2LA state, conditioned on having detected a photon at time $t$, is thus
\begin{eqnarray} 
  |\psi_{{\cal D}}\rangle &=& \langle \text{vac} | \hat{b}
  |\psi(t)\rangle
\nonumber \\ &=&
  \sum_n \eta^*(n) \int d^3x\,
  \psi_n (\vec{x},t) |g \rangle \otimes | \vec{x} \rangle 
\end{eqnarray} 
The normalized post-detection 2LA wavepacket $ \psi_{{\cal D}}(\vec{x}) =
(\langle g|\otimes \langle \vec{x}|)| \psi_{{\cal D}}\rangle$
is therefore given by Eq.~(\ref{eq:detect2})
for detector function 
\begin{eqnarray} 
  {\cal D}(\vec{x}) &=& - \frac{d_{eg}^*}{\hbar} \int_{{\cal C}} 
   \frac{ \text{d}s }{2\pi} \text{e}^{ts}  \sum_n \eta^*(n)
  \frac{E_n^*(\vec{x})}{is-\omega_n}  
\nonumber \\ & & \times
  \left (  i s -\omega_A - \frac{|d_{eg}|^2}{\hbar^2} \sum_m
    \frac{|E_m(\vec{x})|^2}{is-\omega_m}  \right )^{-1}
\label{emDetectorFunction}\end{eqnarray} 
Hence, detecting spontaneously emitted radiation from an atom interferometer
corresponds to a generalized position measurement, whereby the effect
of arbitrary linear optical elements 
only affects the form of the detector function ${\cal D}\left(\vec{x}\right)$.
We remark that in free space this fact can also be explained by the entanglement
between the photonic momentum and the atomic center-of-mass motion due to
momentum conservation~\cite{Len95}.

Our result can be used to resolve unambigously the question whether a 
perfect lens would challenge causality: because a perfect lens can also 
be described as a linear optical device, it can only affect the shape of
${\cal D}\left(\vec{x}\right)$. Hence Inequality~(\ref{eq:WVineq}) is fulfilled,
and a perfect lens would not contradict
quantum mechanics. It simply would allow to increase $W$ at the expense of
reducing $V$.
The reason is that the effect of detecting a photon has a purely local effect and does
not introduce any correlations between different parts of the atomic wave
packets. This is a direct consequence of neglecting the kinetic center-of-mass
energy of the atoms, which is possible because for most systems the electronic
dynamics is fast compared to the motion of the atomic nucleus. 
In free space the effect of the extension of the atomic wavepacket
on spontaneous emission has been discussed in Ref.~\cite{Rzc92}.

\section{Example: diffraction limit and the thin lens}
In this section we apply the formalism developed above to 
a particular physical situation that is related to the
case of the Heisenberg microscope: we consider 
the case that the which way detector is so far away from the 
interferometer that the spontaneous decay of the atom is 
practically completed before the photon enters the detector.
Our assumption corresponds to the far field limit.
If the far field limit is not achieved in an experiment,
full separability of detector and source modes is not
achieved, and a clean signature of complementarity
would then be somewhat masked.
The which way detector consists of a thin conventional lens and the actual detector;
Fig.~\ref{microscSketch} depicts the spatial 
arrangement of the 2LA, lens, and detector.
We will derive expressions for $V,W$ and the uncertainty
of modular momentum and show explicitly how they are affected by the
diffraction limit of the lens.

Under these assumptions the atomic spontaneous decay can be treated as in
free space. The modes of the radiation field introduced in 
Sec.~\ref{Sec:PerfectLensPlusComplementarity} therefore
correspond to plane waves.
Replacing the sum over $n$ in Eq.~(\ref{emDetectorFunction})
by an integral over the wavevector $\vec{k}$ of the modes
we have
\begin{equation} 
  E_{\vec{k}} (\vec{x}) = \sqrt{\frac{\hbar \omega_k}{2\varepsilon_0 (2\pi)^3}}
    \, e^{i \vec{k}\cdot \vec{x}}
\end{equation} 
with the dispersion relation $\omega_k = c |\vec{k}|$. This results in
\footnote{Because all quantities related to complementarity and uncertainty 
are invariant under a rescaling of ${\cal D}(\vec{x})$ we can ignore all
constant prefactors in the derivation.}
\begin{eqnarray} 
  {\cal D}(\vec{x}) &\propto&  \int_{{\cal C}} 
  \text{d}s\,  \text{e}^{ts}  \int d^3k\, \eta^*(\vec{k}) e^{-i \vec{k}\cdot \vec{x}}
  \frac{\sqrt{\omega_k}}{is- \omega_k}  
\nonumber \\ & & \times
  \left (  i s -\omega_A - \frac{|d_{eg}|^2}{2\varepsilon_0 (2\pi)^3 \hbar} \int d^3k' \,
  \frac{\omega_{k'}}{is-\omega_{k'}}  \right )^{-1}
\end{eqnarray} 
In Wigner-Weisskopf approximation \cite{milonni}  we can replace the integral 
over $\vec{k}'$ (including its prefactors) by $\Delta_\text{L} -i \gamma/2$. We absorb
the Lamb shift  $\Delta_\text{L}$
into the definition of the resonance frequency so
\begin{eqnarray} 
  {\cal D}(\vec{x}) &\propto&  \int_{{\cal C}} 
  \text{d}s\,  \text{e}^{ts}  \int d^3k\, \eta^*(\vec{k}) e^{-i \vec{k}\cdot \vec{x}}
  \frac{\sqrt{\omega_k}}{is- \omega_k}  
\nonumber \\ & & \times
  \left (  i s -\omega_A + i \frac{\gamma}{2}  \right )^{-1}
\end{eqnarray} 
Closing the path ${\cal C}$ and using the residue theorem yields
\begin{eqnarray} 
  {\cal D}(\vec{x}) &\propto&  \int d^3k\, \eta^*(\vec{k}) e^{-i \vec{k}\cdot
    \vec{x}}
  \frac{ \sqrt{\omega_k} }{\omega_k -\omega_A +i \gamma/2 }
\nonumber \\ & & \times
  \left (
    e^{-i \omega_k t} - e^{-i \omega_A t -\gamma t/2}
  \right )\; .
\end{eqnarray} 
For sufficiently long times, $\gamma t \gg 1$, the emission process is
completed and the detector function reduces to
\begin{eqnarray} 
  {\cal D}(\vec{x}) &\propto&  \int d^3k\, \eta^*(\vec{k}) e^{-i \vec{k}\cdot
    \vec{x}}
  \frac{ \sqrt{\omega_k} }{\omega_k -\omega_A +i \gamma/2 }
    e^{-i \omega_k t} \; .
\label{DThinLens}\end{eqnarray} 

The detector function ${\cal D}(\vec{x})$ depends on the detection device
through the function $\eta(\vec{k})$. 
After the photon has passed the lens 
it propagates for a certain time until
it reaches the image plane at which the detector is placed.
If the lens is placed at position $z=2f$ the image plane of the
light 
will be at $z=4f$.
To travel a distance $2f$, light propagates for time $t=2f/c$.
Because the 2LA is located close to the origin, the detector should
be in the image plane of the lens at $z=4f$. 
The detector itself is assumed to
respond to photons in a certain spatial mode $ \bar{\eta}(\vec{x})$ with Fourier
transform
\begin{equation} 
  \bar{\eta}(\vec{k}) = \frac{w_\|^{1/2} w_\perp}{\pi^{3/4}}
   \text{e}^{-(w_\perp^2 \vec{k}_\perp^2 + w_{\|}^2 k_z^2) /2} 
       \text{e}^{4 i k_z f} .
\end{equation} 
Here $w_\perp$ and $w_{\|}$ denote the width of the detector mode transverse
to and along the $z$-axis, respectively, and
$\vec{k}_\perp \equiv k_x \vec{e}_x + k_y \vec{e}_y$
In the following we will ignore the degrees of freedom along the
$z$-direction because it is irrelevant for complementarity
of $\hat{p}$ and $\hat{x}$ in the transverse direction.
The lens represents a linear optical device, which generally effects 
a linear transformation of the detector mode of the form
\begin{equation} 
  \eta(\vec{k}) = \int \text{d}^3\vec{k}'\,
M(\vec{k}, \vec{k}')\bar{\eta}(\vec{k}')  
\end{equation} 
For the case of the single conventional thin lens in front of the detector,
the transfer function is
\begin{equation} 
  M(\vec{k}, \vec{k}') \propto  
  \exp \left ( -\frac{1}{2}\frac{(\vec{k}_\perp - \vec{k}_\perp')^2}{
   \frac{1}{L_\perp^2} -i \frac{k_0}{f} } \right ) \; ,
\label{GaussianLens}\end{equation}  
with $f$ the focal length of the lens and $L_\perp$ the radius. 
For $L_\perp \rightarrow \infty$ this expression coincides with 
the usual transfer function for an infinitely wide thin lens.
In a more accurate model for a thin lens its finite size would
be taken into account by a step function
$M(x,x') \propto \theta(L_\perp^2 - x^2 - y^2)$ in position space. 
To simplify the discussion we use instead a model where the finite
size of the lens is taken into account by a Gaussian spatial weight factor
$\propto \exp(-(x^2+y^2)/L_\perp^2)$. 
This procedure generates the term $L_\perp^{-2}$ in 
Eq.~(\ref{GaussianLens}) and leads to
\begin{equation} 
  \eta(\vec{k}) \propto
  \exp \left (
   - \frac{1}{2} \vec{k}_\perp^2 \left ( \frac{1}{L_\perp^2} +
     \frac{1}{w_\perp^2}
     -i \frac{ k_0}{f} \right )^{-1}
  \right )
\end{equation} 
To simplify the discussion of complementarity we ignore the
details of the spontaneous emission process by setting
$\omega_k \approx \omega_A$ in the non-exponential terms of
Eq.~(\ref{DThinLens}). Furthermore, in the spirit of the paraxial
approximation we make the expansion $\omega_k \approx c k_0 +
c \vec{k}_\perp^2 /(2k_0)$ in the exponentials.
The integrand is then a Gaussian and leads to
\begin{equation} 
  {\cal D}(\vec{x}) \propto
  \exp \left (
    - \frac{ x^2+y^2}{2w_\text{eff}^2} + i\frac{k_0}{f} \delta\phi 
      \frac{ x^2+y^2}{2} 
  \right ) .
\label{psib5a}\end{equation} 
For a small width of the detector, 
$w_\text{eff}$
and the phase shift factor
$\delta\phi$ associated with the wavefront are given by
\begin{eqnarray}  
  w_\text{eff}^2 &\approx& \frac{4f^2}{k_0^2 L_\perp^2} + w_\perp^2
\\
  \delta\phi &\approx& \frac{1}{2} + \frac{k_0^4 L_\perp^4}{32 f^4} w_\perp^4\, .
\end{eqnarray} 
This implies that the detector function is diffraction-limited with minimal effective width 
$w_\text{min} =2f/k_0 L_\perp$, 
which corresponds to Heisenberg's and Bohr's
analysis of the Heisenberg microscope:
the resolution limit of a
microscope led them to infer the position uncertainty
$\Delta x_\text{Heis} = \lambda/2 \sin \alpha$ with $\lambda$ the
wavelength and $\alpha$ the opening angle of the microscope's lens.
For $f \gg L_\perp$ we have $L_\perp /f = \tan \alpha \approx \sin \alpha$
and therefore $\Delta x_\text{Heis} = \pi w_\text{min}/2$; the difference
in the numerical prefactor is due to the Gaussian lens approximation that we have used.

\begin{figure} 
\includegraphics[width=7.5cm]{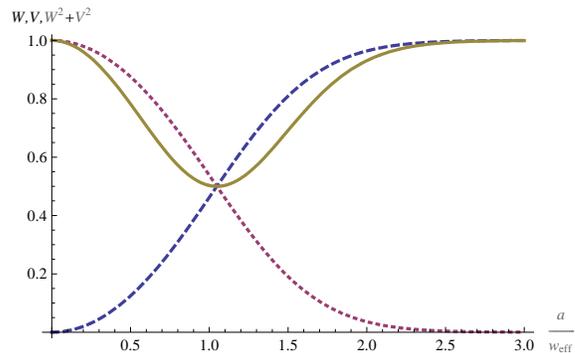} 
\caption{\label{thinLensComplementarity} 
  (Color online) Which way information (dashed line), visibility (dotted line),
 and duality $W^2+V^2$ (solid line) for a thin lens detector and
 well separated wavepackets as a function of the separation $a$
 of the wavepackets in units of the detector resolution $w_\text{eff}$.} 
\end{figure}

In the case that the wave function $\phi\left(\vec{x}\right)$
is a Gaussian with width $ w_\phi \ll w_\text{eff}$ and the two wavepackets
$\phi(x), \phi(x-a)$ are well separated
one finds
\begin{eqnarray} 
  W &=& \tanh \left (\frac{a^2}{2 w_\text{eff}^2}\right )
\\
  V &=& \frac{2}{1+\exp \left ( \frac{a^2}{w_\text{eff}^2}\right )}
\end{eqnarray} 
The exponential damping terms $\exp(-a^2/w_\text{eff}^2)$ reflect the fact that if the
atomic wave function distance $a$ is much larger than the width $w_\text{eff}$
of the detector function, then the photo detection will allow to distinguish the
two wavepackets. It then allows us to gather information 
about $W$ and thus diminish $V$. This behaviour is shown in 
Fig.~\ref{thinLensComplementarity} where
$W$ and $V$ are plotted as a function of the separation between the two wavepackets.
It is apparent that the duality relation is always satisfied. For very small
(very large) separations the inequality is saturated because in these cases 
the photo emission generates no (maximal) which way information, respectively. 

The mean value of modular momentum is
\begin{equation} 
  \langle \psi_{{\cal D}} | \hat{T}_\vec{a} |\psi_{{\cal D}} \rangle  =
  \frac{ e^{-\frac{ a^2 }{ 2 w_\text{eff}^2} }  e^{-i \frac{a^2 k_0}{2f}\delta\phi}
  }{ 1+ e^{-\frac{a^2}{w_\text{eff}^2}} } \; .
\end{equation} 
For large separations of the wavepackets it approaches 0 (completely indefinite 
modular momentum) because in this limit the which way detector 
completely destroys the coherence between the two wavepackets.
The phase factor in $\langle \hat{T}_\vec{a} \rangle$
has the following interpretation:
for large enough detectors $\delta\phi\sim 1$, and the shift in
the phase factor corresponds to $\exp (ia\,\delta p_x/\hbar)$, where
$\delta p_x$ is the momentum difference in the $x$-direction
(transverse to the propagation axis) for photons that arrive at the same
point on the lens but are emitted by different wave functions. This is given
by $\delta p_x = $ (total photon momentum) $\times$ 
(wave function separation)/(propagation length) $= \hbar k_0 a/(2f)$.

\begin{figure} 
\includegraphics[width=7.5cm]{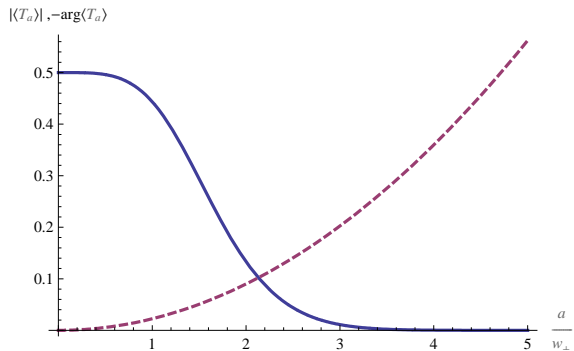} 
\caption{\label{thinLensModularMomentum} 
   (Color online) Modulus (solid line) and phase (dashed line)
 of the mean modular momentum for a thin lens detector and
 well separated wavepackets as a function of the separation $a$
 of the wavepackets in units of the detector resolution $w_\text{eff}$.} 
\end{figure}

In Fig.~\ref{thinLensModularMomentum} we present a numerical example for
the behaviour of $\langle \hat{T}_\vec{a} \rangle$. The parameters chosen
are  $k_0=10^7 \text{m}^{-1}$, $L_\perp = 5$~cm, $f=20$~cm, and $w_\perp = 30\mu$m.
The modulus always less than 1/2 because this is the maximum value for 
$\langle \hat{T}_\vec{a} \rangle$ in the case of well separated wavepackets.
%

\section{Conclusions}
We have analyzed the relation between the duality relation
$W^2+V^2 \leq 1$, which connects which way information $W$
and fringe visibility $V$ in an atom interferometer and
a Heisenberg uncertainty relation between atomic position
and (modular) momentum. A quantitative link between both
concepts can be established by modeling the process of splitting the 
matter beam using the operator $\hat{T}_\vec{a}$ 
of Eq.~(\ref{eq:T}), which spatially shifts the initial wavepacket by
a distance $\vec{a}$. 
This shift operator can also be interpreted as the operator
of the atomic modular momentum. The process
of splitting the atomic beam is therefore naturally connected to a change
in modular momentum. We have shown that this connection allows us
to derive the uncertainty relation from the duality relation 
if $W$ is obtained by a generalized (smeared out) position measurement.

Furthermore we have shown that the detection of spontaneously emitted photons
in an atom interferometer corresponds to a generalized
position measurement, provided the detection device can be described 
using lossless linear optical elements and projection measurements.
Because the duality relation holds regardless of the specific nature
of the detection device, the complementarity principle of quantum
mechanics holds regardless of the quality of the detection device in use.
Complementarity is therefore not affected by superresolving optical
devices or perfect lenses based on meta-materials; such optical elements
can only affect the amount of which way information that can be
gathered, but not the duality relation.

\emph{Acknowledgments.--} We thank A.~Lvovsky for
helpful discussions. This work has been supported
by iCORE, NSERC, CIFAR, MITACS, QuantumWorks, the UK Engineering and Physical Sciences 
Research Council IRC in Quantum Information Processing, and the
European Union networks CONQUEST and SCALA.


\begin{thebibliography}{99}
\bibitem {San89} B. C. Sanders and G. J. Milburn, \pra \textbf{39}, 694
  (1989).
  \bibitem {Scu91} M. O. Scully, B.-G. Englert, and H. Walther, 
  Nature (Lond.) \textbf{351}, 111 (1991).
\bibitem{Jae95} G.~Jaeger, A.~Shimony, and L.~Vaidman, Phys.~Rev.~A {\bf 51},
  54 (1995).
\bibitem {Eng96} B.-G.~Englert, \prl \textbf{77}, 2154 (1996).
\bibitem{Eng00}  B.-G.~Englert and J.~A.~Bergou, Opt.~Commun.~{\bf 179}, 337 (2000).
\bibitem {Wis97}{H. M. Wiseman, Harrison, F.~E., Collett, 
    M.~J., Tan, S.~M., Walls, D.~F. \& Killip, R.~B., 
        {\em et al.}, \pra \textbf{56}, 55 (1997).}
\bibitem {Due00} S. D\"urr and G. Rempe, \emph{Am. J. Phys.} \textbf{68}, 1021 (2000).
\bibitem {Aha04} Y. Aharonov and D. Rohrlich,
  \emph{ Quantum Paradoxes: Quantum Theory for the Perplexed}
  (Wiley-VCH, Weinheim, 2004).
\bibitem {Woo79} W. Wootters and W. H. Zurek, \prd \textbf{19}, 473 (1979).
\bibitem {Sac06} M.\ F.~Sacchi, \prl \textbf{96}, 220502 (2006).
\bibitem {Hei29} W. Heisenberg, Z.~Phys. \textbf{43}, 172 (1927).
\bibitem{storey95} E. P. Storey, S. M. Tan, M. Collett and D. F. Walls,
  Nature {\bf 367}, 626 (1994).
\bibitem{Mir07} R.~Mir, J.~S.~Lundeen, M.~W.~Mitchell, A.~M.~Steinberg,
  J.~L.~Garretson,
  and H.~M.~Wiseman, New J.~Phys. {\bf 9}, 287 (2007).
\bibitem{Aha88} Y.~Aharonov, D.~Z.~Albert, and L.~Vaidman,
  Phys.~Rev.~Lett.~{\bf 60}, 1351 (1988).
\bibitem{Dur98} S.~D\"urr, T.~Nonn and G.~Rempe, Nature {\bf 395}, 33 (1998).
\bibitem{Wis03} H.~M.~Wiseman, Phys.~Lett.~A {\bf 311}, 285 (2003).
\bibitem{Bus06} P.~Busch and C.~Shilladay, Phys.~Rep.~{\bf 435}, p. 1
  (2006).
\bibitem{Bus07} P.~Busch, T.~Heinonen, and P.~Lahti, Phys.~Rep.~{\bf 452}, p. 155
  (2007).
\bibitem{Roy78} C.~Roychoudhuri, Found.~Phys.~{\bf 8}, 845 (1978).
\bibitem {Ber03} P. R. Berman, quant-ph/0309196 (2003).
\bibitem {Pen00} J. B. Pendry, \prl \textbf{85}, 3966 (2000).
\bibitem {Smi04} D. R. Smith, J. B. Pendry, and M. C. K. Wiltshire, Science
  \textbf{305}, 788 (2004).
\bibitem {Boh29} Note added in proof in Ref.~\cite{Hei29}; 
  M.~Jammer, {\em The Philosophy of Quantum Mechanics} (John Wiley, New York,
  1974), Chapter 3.
\bibitem{lehoucq96:_comput} R.~B.~Lehoucq, ACM Trans.~Math.~Softw.~{\bf 22},
  393 (1996).
\bibitem {Car91} O. Carnal and J. Mlynek, \prl \textbf{66}, 2689 (1991).
\bibitem {Kas92} M. Kasevich and S. Chu, \prl \textbf{67}, 181 (1991).
\bibitem{Sto94} E.~P.~Storey, S.~M.~Tan, M.~J.~Collett, and D.~F.~Walls, Nature
  (London) {\bf 367}, 626 (1994).
\bibitem{POVM} I.~D.~Ivanovic, Phys.~Lett.~A {\bf 123}, 257 (1987);
    D.~Dieks, Phys.~Lett.~A {\bf 126}, 303 (1988);
    A.~Peres, Phys.~Lett.~A {\bf 128}, 19 (1988). 
\bibitem{Kni98} P.~L.~Knight, Nature {\bf 395}, 12 (1998).
\bibitem {Hol84}{A. S. Holevo, in 
  \emph{Quantum Probability and Applications to the Quantum Theory 
  of Irreversible Processes}, L. Accardi, A. Frigerio, and V. Gorini, eds., 
  Lecture Notes in Math.~Vol.~1055 (Springer-Verlag, Berlin, 1984), p. 153.}
\bibitem {Len95} M. S. Chapman, T. D. Hammond, A. Lenef, 
  J. Schmiedmayer, R. A. Rubenstein, E. Smith, and D. E. Pritchard, 
        {\em et al.}, 
        \prl \textbf{75}, 3783 (1995).
\bibitem{Rzc92}   K. Rz\c{a}zewski and W. Zakowicz, 
	Journ. Phys. B {\bf 25}, L319 (1992).
\bibitem{milonni} P.~W.~Milonni, {\em The quantum vacuum} (Academic Press,
  Boston, 1994).
\end{thebibliography}
\end{document}